\documentstyle[aps,prl,psfig,twocolumn]{revtex}

\mathchardef\gacc="328

\begin{document}
\draft
\twocolumn[\hsize\textwidth\columnwidth\hsize\csname
@twocolumnfalse\endcsname

\title{Dynamical phases and hysteresis in a simple one-lane traffic model}
\author{Micka{\"e}l Antoni$^{1}$,
Raffaele Cafiero$^2$}

\address{$^1$ Thermodynamique et mod\'elisation des milieux hors d'\'equilibre,
Laboratoire de Chimie des Syst\`emes Complexes,
Universit\'e d'Aix-Marseille III,
Av. Escadrille Normandie-Niemen,
F-13397 Marseille, France}

\address{$^2$ PMMH, Ecole Sup\'erieure de Physique et de
Chimie Industrielles, 10, rue Vauquelin, F-75231 Paris, France}
\date{\today}

\maketitle

\begin{abstract}
A two parameter model for single lane car-following is introduced and its
equilibrium and non-equilibrium properties are studied. Despite its simplicity,
 this model exhibits a rich phenomenology, analogous to that observed in real traffic,
like transitions between different dynamical regimes and hysteresis in
the fundamental flux-density diagram. We show that traffic jams can
 spontaneously
appear in clustered-like structures. In the jammed phase, we observe a slow
 relaxation
phenomenon ruled by the outgoing car flux that determines the hysteretic
dependence of the fundamental flux-density diagram. Coexisting phase regimes are
 also
evidenced so as propagating or stationary density waves. The model can be easily
 calibrated to
reproduce experimental observations.
\end{abstract}


\narrowtext
\vskip2pc]

The occurrence of traffic jams or small traffic congestions without obvious
reasons are common effects that almost every driver has once experienced.
These effects are typical signatures of the complex behavior of traffic
flows and their study is of practical interest in the context of
traffic control. The evolution of such spontaneous time-space structures has
 long
attracted attention in the understanding of non-equilibrium properties of
externally driven many-body systems. Sophisticated concepts were in
particular developed to study critical phenomena and successfully applied
to the description of phase transitions in traffic flow models
 \cite{kern1,krau1}.

Real traffic exhibits a very rich variety of phenomena. For this reason, the
 most
adapted techniques for efficient traffic control are still debated. Experimental
investigations on highways revealed that traffic can exhibit well identified
dynamical regimes that depend on the external car flux and on the car density
\cite{kern2,kern3,kern4}. Phase transitions occur in traffic flows when the
 vehicle
density exceeds a critical threshold. Below this threshold, traffic is free.
 Beyond it, vehicles either briefly slow down due
to high density traffic or stop in a jam. Jams can appear without
obvious reasons, they can merge and extend to large scales. The outcome of
experimental measurements were moreover shown to be strongly influenced by the
traffic behavior near the measurement site. Near ramps
for example, peculiar behaviors such as avalanches or oscillations
are known to take place \cite{lee1}. Traffic flow results are thus far
from universal \cite{nage2}.

Planning and optimizing real traffic flow has motivated many theoretical
 approaches,
ranging from cellular automata models \cite{nage1} to coupled maps
\cite{tada1} and from hydrodynamics \cite{lee1,naga2,kern5,helb2} to kinetic
 theory
\cite{kern1,trei1,helb3,ben1}. Theoretical results depend on the
drivers behavior. In deterministic models for instance non-linearities 
induce the jamming transition and the jammed phase
persists in time \cite{band1} while for stochastic models a jamming transition
 occurs
due to the intrinsic noise and no jam persists for ever \cite{nage3}.
Most of the theoretical models
are based on heuristic arguments and their parameters have to be calibrated using
experimental outputs.

With the aim of retrieving experimental properties of traffic, we focus
on a simple $2$-parameters car following model. We consider a one lane street of
length $L=10$ km. Vehicles on this street are all identical
with size $d_{car}=5$ meters
\cite{helb2} and indexed such that car number $i+1$ precedes car $i$. They
 cannot
overtake and, due to periodic boundary conditions the vehicle ahead car $1$ is
 $N$.
When approaching a slower car the driver slows down. The velocity of the front
car remains unaffected.
Car $i$ has position $x_i(t)$ and velocity $v_i(t) \le v_{max}$
at time $t$, where $v_{max}$ is the maximal legal velocity. We assume that each
driver determines his velocity in function of the headway to the front car. This
is based on safety requirements and on a simple empirical rule. Traffic
security experts indeed often teach to beginners that driving conditions are
 safe at
velocity $v_i$ if the distance $d_i$ with the vehicle ahead is such that
$d_i > v_i$, where $d_i$ is estimated in meters and $v_i$ expressed in km/h.
This means that at velocity $100$ km/h a minimal headway security
distance of $100$ m is required. When following this rule, drivers have $T=3.6$
seconds to react and avoid collisions. This time is
independent of the velocity \cite{may1} and is supposed
to include the front
car stimulus, braking distance and drivers concentration level \cite{priv1}.
The velocity law for car number $i$ at time $t$ is
given by~:

\begin{tabular}{ccc}

\begin{minipage} [c] {1cm}
\begin{displaymath}
v_i(t) =
\end{displaymath}
\end{minipage}

&

\begin{minipage} [c] {0.5cm}
\begin{displaymath}
\gacc
\end{displaymath}
\end{minipage}

&

\begin{minipage} [c] {2.5cm}
\begin{eqnarray}
&v_{max} \quad {\rm for}& \quad d_i(t) \ge 25 d_{car}
\label{mod1}
\\
&d_i(t)  \quad {\rm for}& \quad d_{car} \le d_i(t) < 25 d_{car}
\label{mod2} \nonumber
\end{eqnarray}

\end{minipage}

\end{tabular}

\noindent
where $d_i(t)=mod(x_{i+1}(t)-x_i(t),L)$ and $v_i(t)$ is expressed
in km/h \cite{help1}. $d_{car}$ and $v_{max}$ are the two parameters
of this model. For the latter to be consistent with experimental observations,
we set $v_{max}=125$~km/h \cite{kern3}. The prefactor $25$ in the first equation
is then a consequence of the empirical driving rule. Indeed, in order to proceed
safely, a driver at velocity $v_{max}$ has to slow down when the headway
 distance
becomes smaller than $125$~m~$=25d_{car}$.
Although very simple, we show herein that simulations of Eqs. (\ref{mod1})
reproduce spatio-temporal observations of experimental traffic flows. For the
 time evolution
of the position of each car we use a synchronous updating based on the Euler
 scheme
$x_i(t+\Delta t)=x_i(t)+v_i(t)\Delta t + O(\Delta t)^2$ with time step
$\Delta t=0.1$~s. Due to this time discretization and in order to avoid
artificial car crash we have to constraint numerically $v(i)=0$
when $d(i) < d_{car}$. This introduces a small dynamical noise that
slightly perturbs our simulations. We checked however that the main results
presented herein still hold for $\Delta t=0.01$~s.

We study model (\ref{mod1})
under driven external car flux, {\it {ie.}} we allow the average car
density on the street $\rho$ to vary in time due to in-going (resp. out-going)
cars through an on-ramp (resp. off-ramp) located at position $x=0$ km
(resp. $x=L=10$ km).
To check the validity of this model, we first pay attention to the experimental
 observations
of reference \cite{kern2}. For given position $x$ on the street, we consider the
 one
minute averaged velocity $v_x$ of the cars crossing $x$ and $\rho_x$ the one
 minute averaged density
in a $1$ km band centered on $x$. We deal with one realization of
the system and we model the external car flux by a simple superposition of
three Maxwellian functions. For this, we constraint the time evolution of the
 car density
on the whole street to be $\rho(t)=\rho_0 \sum_{i=1,2,3} \exp[-((t-t_i)/2)^2]$,
with $t_1$=7:00, $t_2$=12:00 and $t_3$=17:00 and where $t$ is expressed in
 hours.
Using the experimental results of \cite{kern2} we set $\rho_0 \sim 50$ cars/km
and $t_i$'s are supposed to be rush periods. Practically, we estimate the
density $\rho(t)$
every $10$ seconds and, according to its value, either keep the total number
of cars to the same value or add/remove a single car from the street. This means
that the maximal car flow is $360$ cars/h which is consistent with constant
external flux models \cite{lee1} but below experimental observations
 \cite{kern4}.

\begin{figure} [h]

\begin{tabular}{cc}
\hglue-0.7cm
\psfig{figure=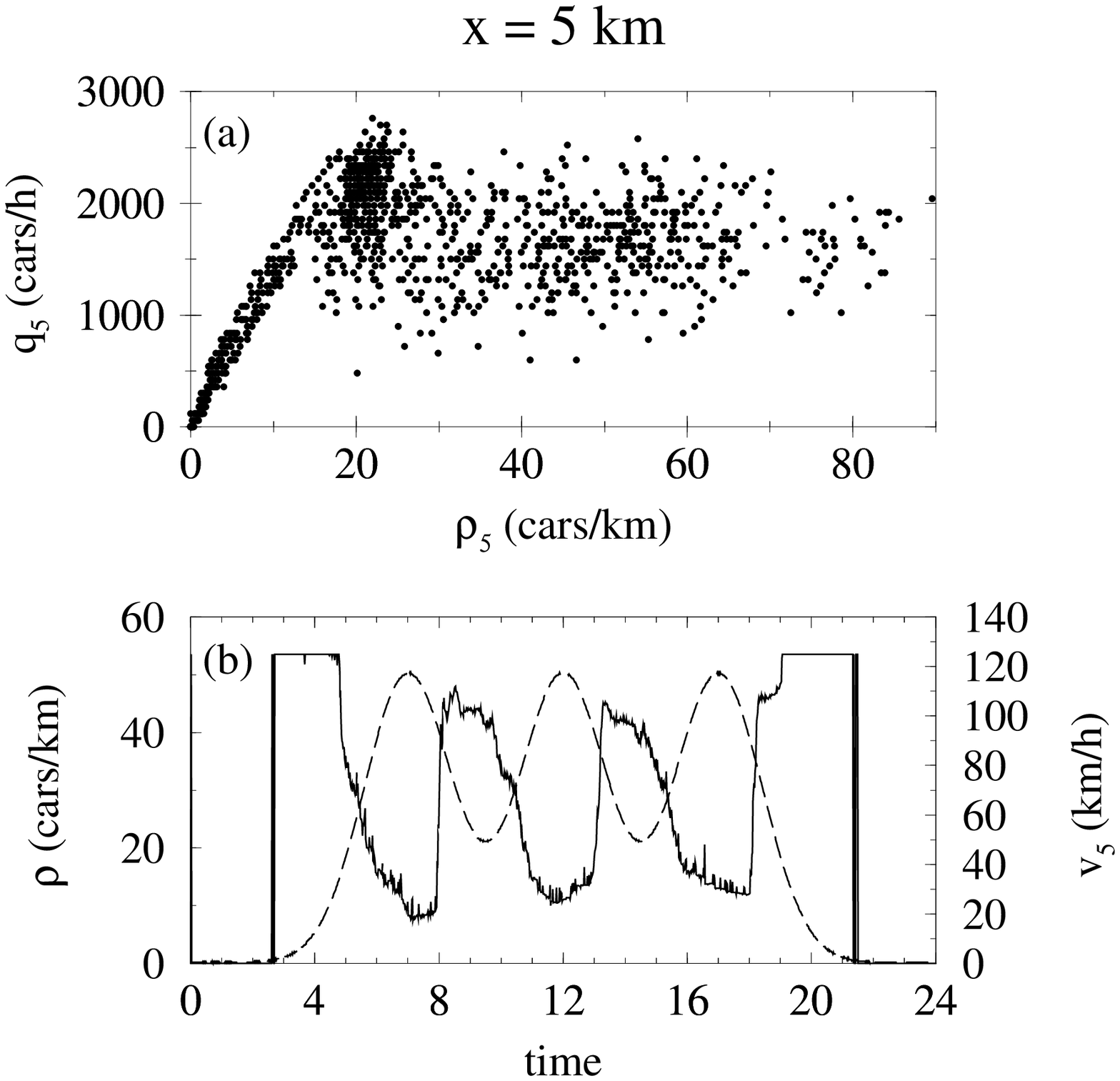,angle=270,width=4.5cm,height=5.6cm}&
\hglue-0.cm
\psfig{figure=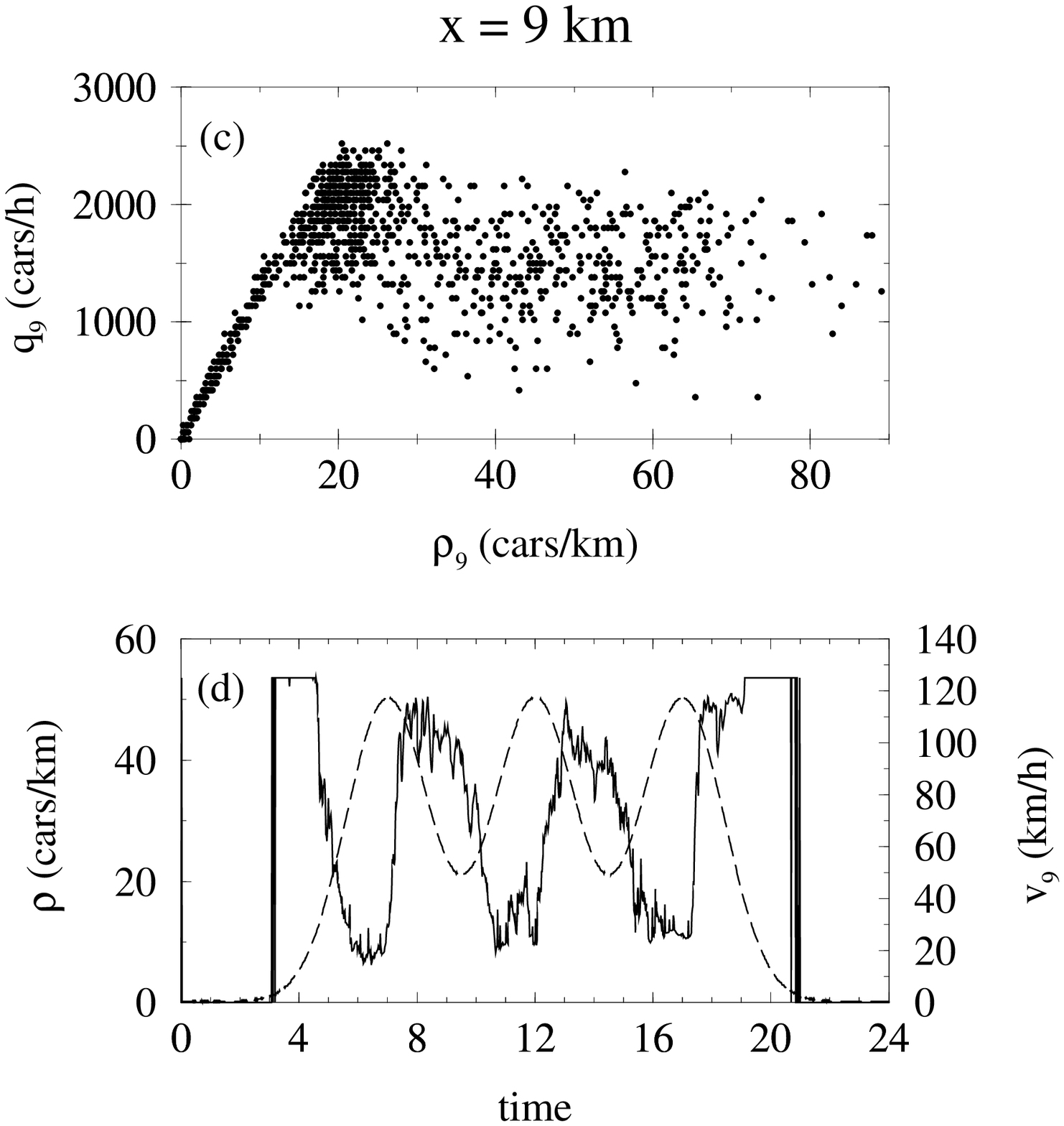,angle=270,width=4.5cm,height=5.6cm}
\end{tabular}

\caption{Flux-density ($q_x,\rho_x$) diagram (upper graphs) and average speed
 $v_x$
(lower graphs full line) of the cars at crossing sections $x = 5$ km
(left column) and $x = 9$ km (right column) during one minute intervals and for
$24$ hours. The dashed
line in (b) and (d) corresponds to the time dependence of the total car
density $\rho(t)$ on the street.}
\label{fig1}
\end{figure}

\begin{figure} [h]
\centerline{\psfig{figure=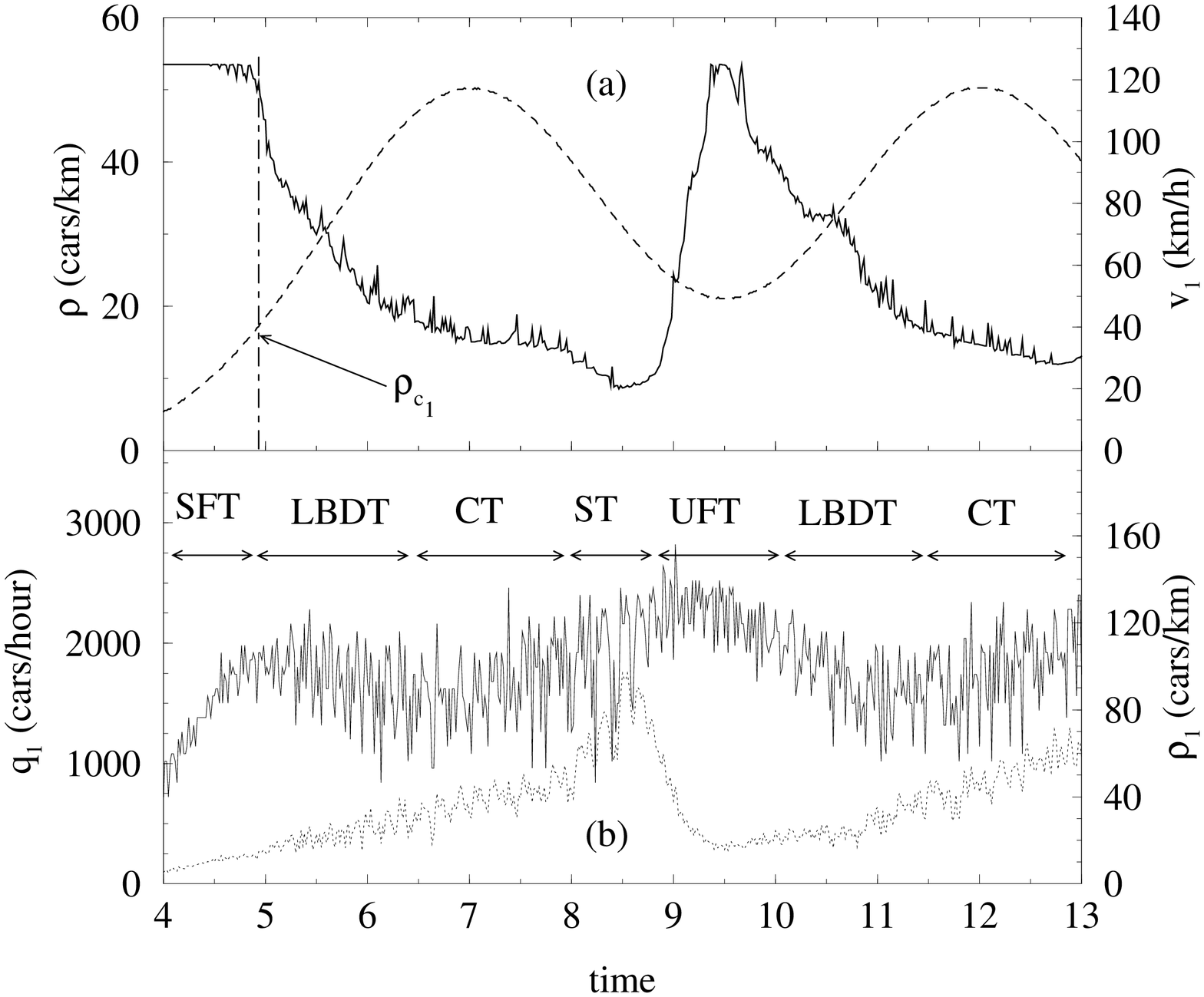,angle=270,width=8.0cm,height=4.6cm}}
\caption{Detail of the time dependence of the average velocity $v_1$ (a), flux
 $q_1$
and density $\rho_1$ (b) at cross section $x=$~1~km. Legends and simulation
 conditions
are the same than in Fig. \protect{\ref{fig1}}. The vertical dot-dashed line
in (a) indicates the moment where the phase transition between free traffic flow
 and
congested flow with critical density $\rho_{c_1} \sim 15$ cars/km. The dotted
 line
in (a) corresponds to the time dependence of $\rho_1$.In (b) we
show the succession between $SFT$, $LBDT$, $CT$, $ST$ and $UFT$.}
\label{fig2}
\end{figure}

In Fig. \ref{fig1} we display the local flux-density $(q_x,\rho_x)$
fundamental diagram and the time evolution of velocity $v_x$. We consider two
cross sections~: The first is situated at position $x=5$ km and the second near
the off-ramp at $x=9$ km. The $(q_x,\rho_x)$ representation of Figs. \ref{fig2}
look very similar to the one obtained in \cite{kern3}.
The numerical points tend however to accumulate in the neighborhood of
$\rho=30$ cars/km. This effect is a consequence of the simplicity of the time
dependence of $\rho(t)$ that tends to leave the system longer near $\rho = 25$
 cars/km.
It can be removed by adding small random fluctuations to $\rho(t)$.

When comparing $v_5$ and $v_9$, we observe larger fluctuations in the
time behavior of $v_9$ (see full line in Fig. \ref{fig2}(b) and (d)).
They appear in time periods where the variations of $\rho(t)$
are fast ({\it{i.e.}} when the external flux is large).
These fluctuations are of particular interest near on-ramps where they can be
 seen as
signature of avalanche-like effects that take place when cars arrive in a
yet ``busy'' highway. 

\noindent
The effect of ramps was investigated both experimentally
\cite{kern3} and theoretically \cite{lee1,kern5}.
This local breakdown effect triggers jams and  remains even if fluctuations in
the external flow are negligible.
Time evolution of $v_1$ and $q_1$ is shown in Fig. \ref{fig2} between time 4:00
 and
13:00. Both figures exhibit five different dynamical regimes: stable free
 traffic
($SFT$ for $t$$\in$[4:00,5:00]), local break down traffic ($LBDT$ for
 $t$$\in$[5:00,6:30]),
clustered traffic ($CT$ for $t$$\in$[6:30,8:00]),
synchronized traffic ($ST$ for $t$$\in$[8:00,9:00]) and
unstable free traffic ($UFT$ for $t$$ \in$ [9:00,11:30]). In order to identify
 these
regimes, we applied a noise to $\rho(t)$ and observed that all the regimes but
$UFT$ were not significantely modified. $UFT$ turns out to be sensitive to
perturbations and disapears for too large noise amplitudes \cite{trei1}. $LBDT$
 on the
other hand appears
when the density grows beyond $\rho_{c_1} \sim 15$ cars/km and is charaterised
by an avalanche-like deacreasing velocity $v_1$ for increasing density
 \cite{kern5}.
This regime is located in the vicinity of the ramps and triggers complex
 non-homogeneous
traffic structures that evolve over the whole street (see Fig. \ref{fig4}).
$CT$ consists in small jammed islands that appear away from ramps, persist in
 time
and can be compared to droplets that form in supercooled gas.
$ST$ phase appears in highway measurements
when a similar average car velocity is observed on all the lanes of the
highway and is mostly localized in the neighborhood of the ramps. One line
 descriptions
were however successfully used to study this regime \cite{lee1,helb3}. $ST$ is
 characterized by a low average
velocity but a rather high value of the local car flux $q_1$ when compared to
 the one
of the jammed traffic phase ($JT$). This latter phase does not appear in the
 previous
figures since $\rho_0$ is too small. $JT$ is a stable phase (any perturbation
 fade
as time evolves) and is characterized by densely packed queues of cars that can
extend in a single jam over the whole street.

\begin{figure} [h]
\psfig{figure=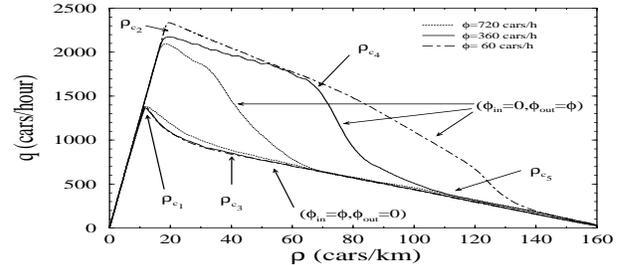,angle=270,width=8.0cm,height=3.6cm}
\caption{Flux-density fundamental diagram averaged over $100$ realizations
for several values of the in/out car flow $\phi$. $\rho_{c_2} \sim 20$ cars/km
determines the threshold beyond which $UFT$ appears. The lower branch (resp.
 upper
branches) of the hysteresis loop is (resp. are) generated during the in-flow
(resp. out-flow) period. $\rho_{c_3} \sim 40$, $\rho_{c_4} \sim 70$
and $\rho_{c_5} \sim 110$ cars/km are defined in the text and refer to
 simulations
with $\phi = 360$ cars/h (full lines). For this value of $\phi$
congested traffic appears when $\rho \in [\rho_{c_2}, \rho_{c_5}]$ whereas $JT$
 appears for
$\rho > \rho_{c_5}$ cars/km.}
\label{fig3}
\end{figure}

Many features of traffic flows can be compared to fluid dynamics. For instance
 $JT$
can be seen as a compressible liquid without coexisting vapor whereas $SFT$
can be trivially compared to a dilute gas \cite{krau1}. Consequently, a
 criterion for
model (\ref{mod1}) to be realistic constraints the transtion between $JT$
and $SFT$ to be histeretic as the liquid-gas transition. This prompts us to
 rapidly
focus on the ensemble averaged total flux $q$ and total density $\rho$.
For each realization of the system, we perform a simulation that starts with
 $\rho=1$
car/km, with a zero out flow $\phi_{out}= 0$ of cars and a constant in flow
$\phi_{in}=\phi$ ({\it {in-flow period}}). Cars are introduced on the street one
 by one, to keep the system as close as possible to its stationary state, and
$q$ is averaged over the time running between two introductions. When density
$\rho_{max} \sim 160$ cars/km is reached, the street is uniformly jammed.
At this density, in-flow is interrupted and we start to remove
cars one by one setting $\phi_{in}=0$ and $\phi_{out}=\phi$. We proceed this way
down to density $\rho=1$ car/km ({\it {out-flow period}}).
This procedure is repeated over $100$ realizations. Fig. \ref{fig3} displays
the ensemble averaged ($q,\rho$) fundamental diagram for several values of
$\phi$. The $SFT$ and the $UFT$ regimes clearly show up for $\rho < \rho_{c_1}$
and $\rho \in [\rho_{c_1}, \rho_{c_2}]$ respectively, with $\rho_{c_2} \sim 20$
 cars/km.
Both regimes do not strongly depend on the value of $\phi$. For $\rho >
 \rho_{c_1}$
we also retrieve the characterisitic hysteresis loop of traffic behavior.
The lower branch
corresponds to in-flow periods. It is independent of
the in-flow rate $\phi_{in}$. This is clearly not the case
for out-flow periods as shown by the existence of several upper branches in
Fig. \ref{fig3}. A direct consequence is that $JT$ occurs beyond a threshold
density $\rho_{c_5}$ that itself depends on the out-flow rate $\phi_{out}$,
where $\rho_{c_5}$ is defined as the density $\rho > \rho_{c_2}$ for which
 the two branches of the
histeresis loop differ less than $10 \%$. Hence, the global stability of
model (\ref{mod1}) is only determined by the out-going car flow $\phi_{out}$.
This property remains true even when modifying the model \cite{com1}. 

For $SFT$, $UFT$ and $JT$ regimes
the corresponding spatial distribution
of the cars is uniform over the whole street. Since we focus on ensemble
 averaged
quantities, only limited informations can be obtained from Fig.\ref{fig3} on the
complex space-time patterns of congested traffic. It is however interesting
to note that for $\rho > \rho_{c_3} \sim 40$ cars/km the in-flow branch is
almost linear. This behavior is similar to
the coexisting regime in fluid dynamics with the fluid and its gas in thermal
 equilibrium. Fig. \ref{fig3} suggests that such a picture also holds
for traffic flows although these systems are far from equilibrium. The
 explanation
is that in this density range the system decomposes in two phases:
A low density one ($\rho_{c_3} < \rho < \rho_{c_5}$)
that we identify as the $CT$ and a high density one ($\rho > \rho_{c_5}$)
corresponding to $JT$. This is illustrated in Fig. \ref{fig4} where we
show the space-time density contour plot. Clustered structures of $CT$
show up in Fig. \ref{fig4} when $\rho_{c_3} < \rho < \rho_{c_5}$. As the density
inside clusters is close to the critical value $\rho_{c_5}$ this situation can
be compared to the small droplets that form in a gas near condensation.
As new cars arrive on the road, the $JT$ first appears at time
$t$=1:7 near the ramp and coexists with $CT$. As time evolves, due to increasing
$\rho$, $JT$ extends backwards and finally overcomes $CT$ (at time 3:00 where
$\rho \sim \rho_{c_5}$). Forward propagating density waves are evident
in  Fig. \ref{fig4}. The contour plot for $\rho > \rho_{c_5}$
in Fig.\ref{fig4} shows complex space time structure and an almost uniform
$JT$ phase that extends over the complete street when $\rho \sim \rho_{max}$.

Nucleated structures of $CT$ also show up in the out-flow period when
 $t$$\in$[6:7]
($\rho_{c_4} <\rho < \rho_{c_5}$). But in this time range, it coexists with two
 quiet
well separated $JT$ and $SFT$ phases. For longer times, when $\rho_{c_2} <\rho <
 \rho_{c_4}$,
a new separated phase between $JT$ and $SFT$ appears whit two density fronts.
As $\rho$ decreases with time, the in-flow of
cars is no more sufficient to maintain $JT$ that progressively disapears. In the
down stream direction the front is standing while in the upstream one, the front
propagates due to the 'evaporation' of the $JT$ phase. From these results,
 $\rho_{c_4}$
turns out to play the role of a physical threshold between a triphasic
 $JT$-$CT$-$SFT$
situation and a situation where $JT$ and $SFT$ coexist in a simple space-time
shrinking structure.

\vglue-2.0cm
\begin{tabular} {cc}
\begin{minipage}{2.7cm}
\vglue0.2cm
\begin{figure} [h]
\hglue-0.1cm\psfig{figure=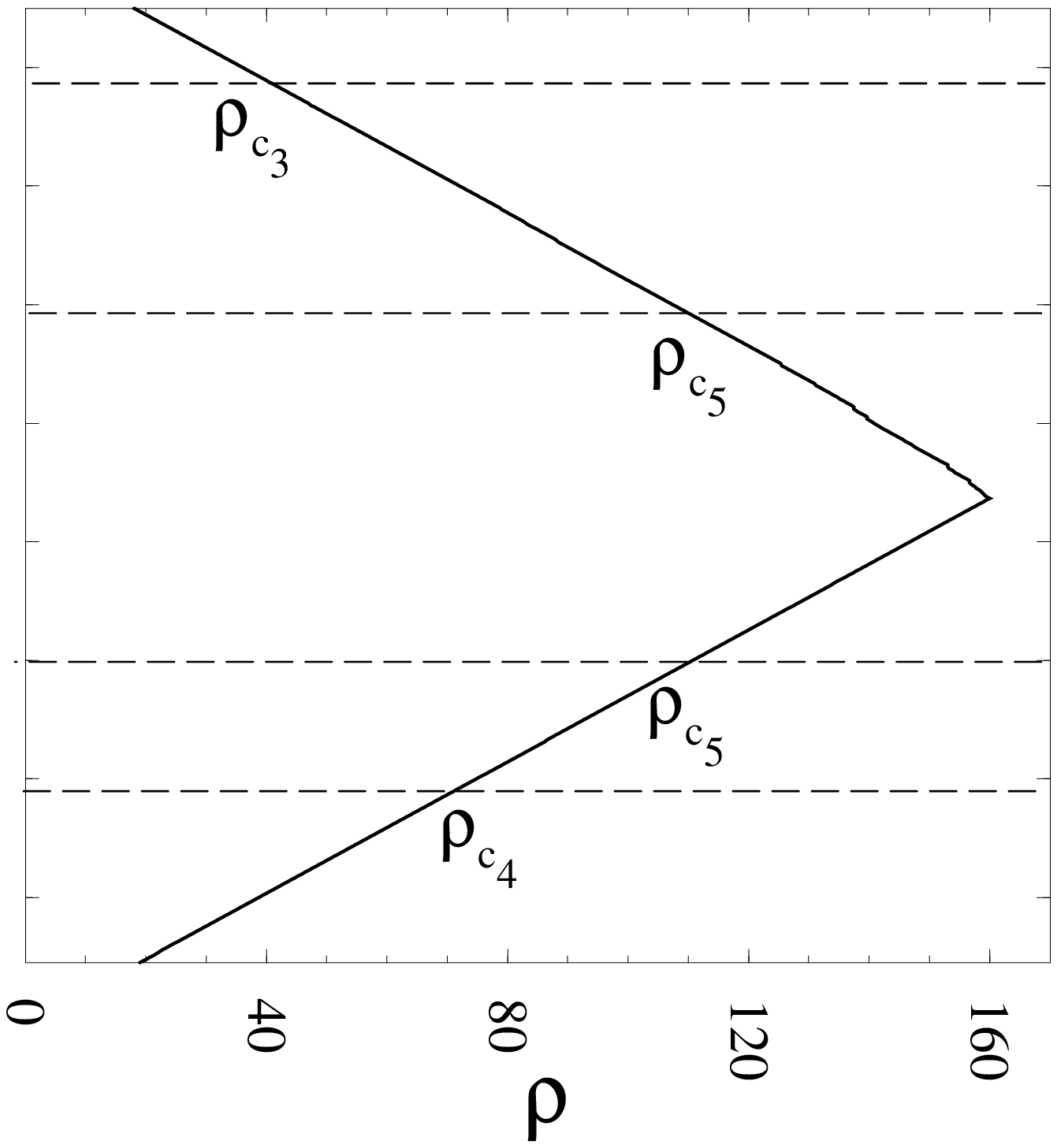,angle=180,width=2.15cm,height=7.2cm}
\end{figure}
\end{minipage}
&
\begin{minipage}{7cm}
\vglue-0.8cm
\begin{figure} [h]
\hglue-1.0cm\psfig{figure=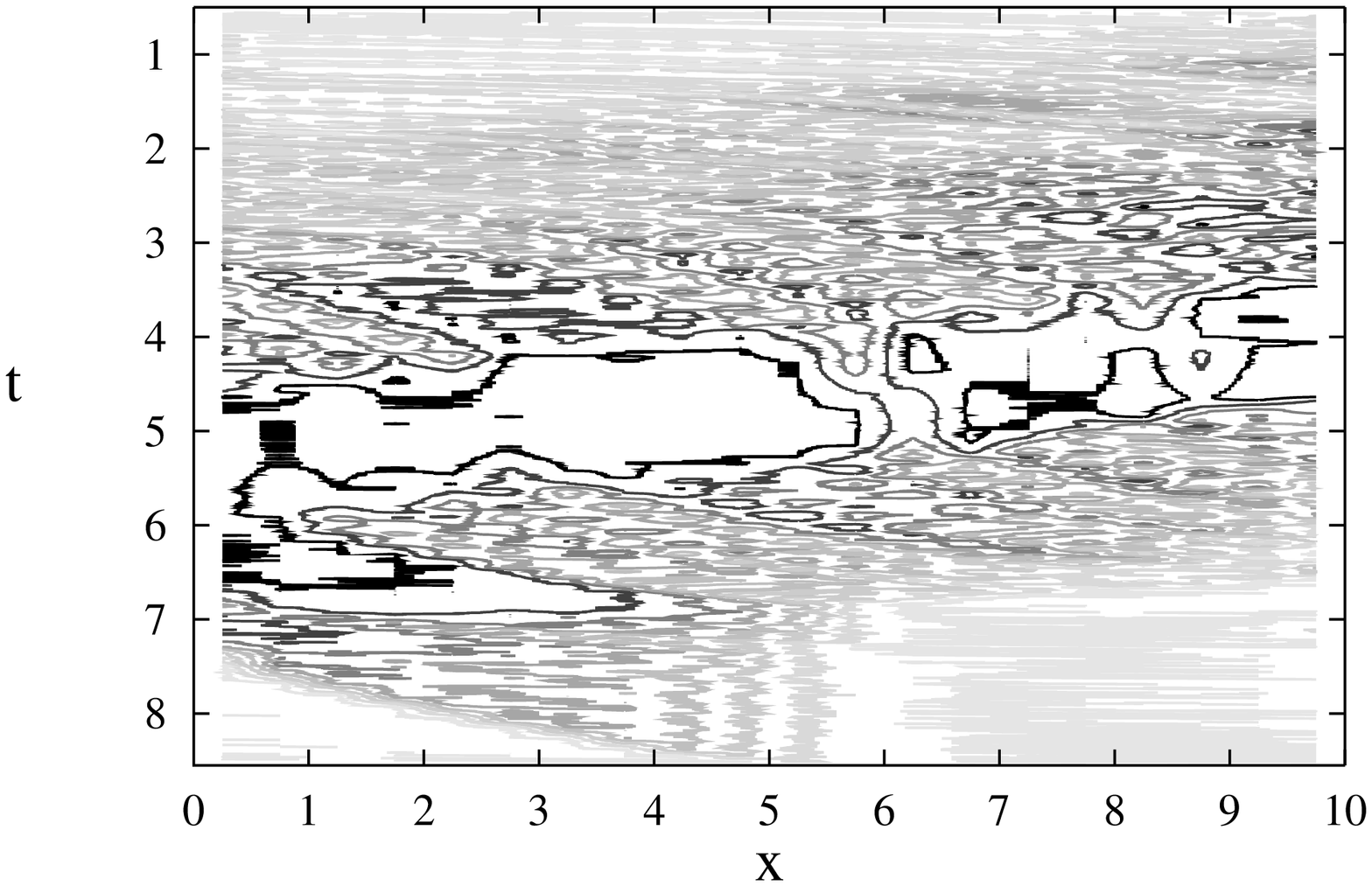,angle=-90,width=7.47cm,height=11.62cm}
\end{figure}
\end{minipage}

\end{tabular}

\vglue-8.25cm
\begin{figure}
\psfig{figure=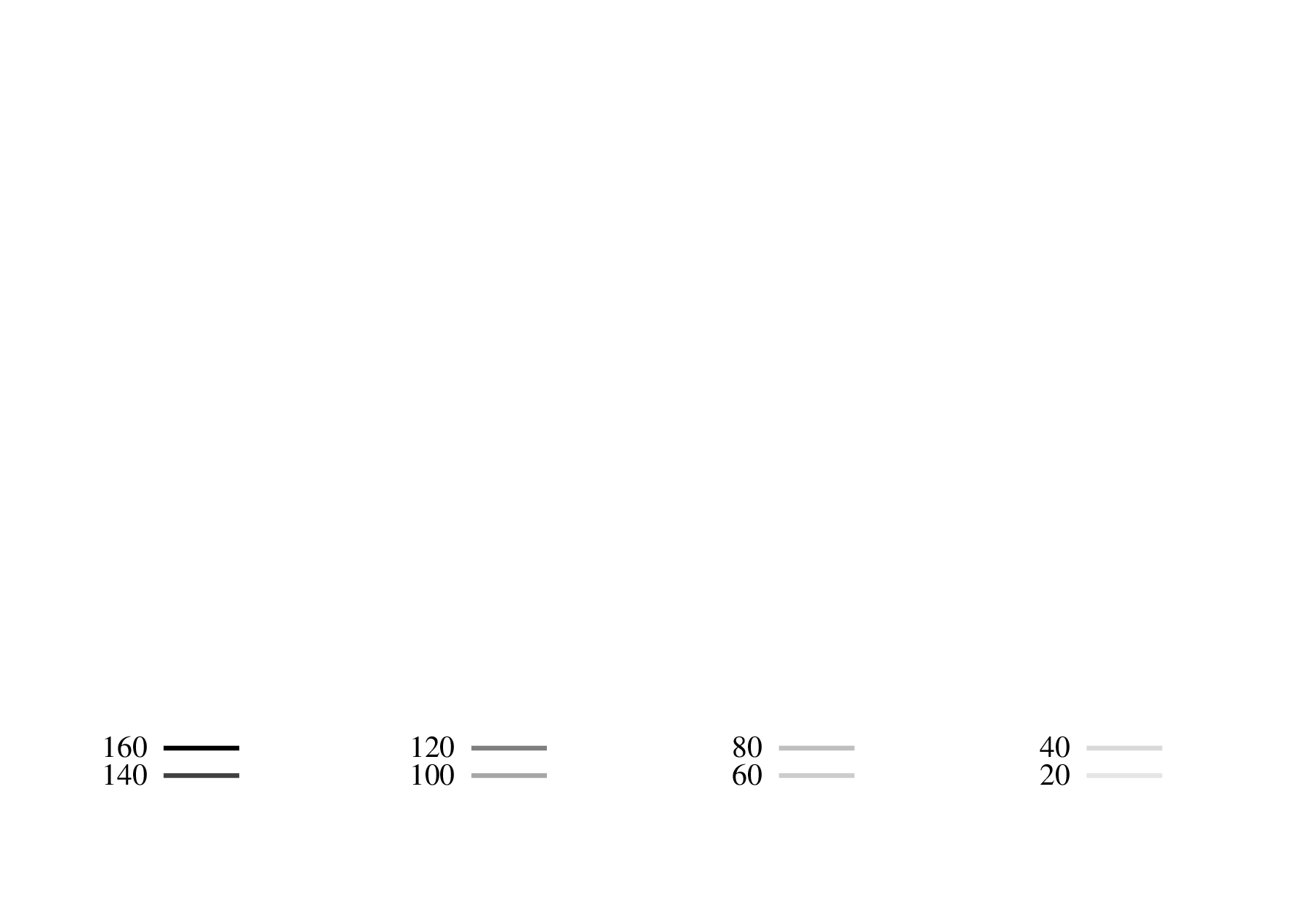,angle=-90,width=7.47cm,height=7.47cm}
\end{figure}
\vglue-1.6cm
\begin{figure}
\caption{Time evolution of the total density $\rho$ (left) and space-time
density contour plot (right) for the run of Fig. {\protect \ref{fig3}}
with $\phi=360$ cars/h, during in flow $t<$4:30 and out-flow $t>$4:30 periods.
In these figures we restricted ourselves to the time period where $\rho \ge
 \rho_{c_2}$.
In the left graph, the horizontal lines indicate the time at which the
critical density $\rho_{c_i}$, $i=3,4,5$, is reached. The street is divided in
 $20$ sections
of $500$ meters and in each of them the density $\rho$ is evaluated every
 minute.
The legend of colors are given in cars/km.}
\label{fig4}
\end{figure}

We considered herein a simple $2$-parameters microscopic
car-following model. When driven
far from equilibrium by an external car flux, this model reproduces
the main characteristics of experimental fundamental diagram. Near ramps,
we retrieve avanlanche-like effects so as synchronized traffic.
The ensemble averaged fundamental diagram exhibits
the traditional histeresis loop of traffic models. This a signature
of coexisting phases and we showed that among these phases two regimes
have to be distinguished: a clustered phase where the system develops complex
space time patterns and a regime where phases are separated by well identified
density fronts.

The authors are grateful to the Max-Planck-Institute for Physics of Complex
 Systems of
Dresden (Germany) for hospitality
and computer power.
M. A. thanks Dr. J. Kister for determinant help during his installation in the
University of Aix-Marseille III, Prof. A. Steinchen and Dr. B. Roux for
 fruitfull discussions, and C. Albertini for computational support. R. C.
 acknowledges financial support under the European network project FMRXCT980183.


\begin{references}
\bibitem{priv1} Similar driving
rules are briefly mentionned in \protect{\cite{nage2}} and
 \protect{\cite{trei1}}.
\bibitem{help1} Velocity is given by $v_i=d_i/T$ in m/s. As $T=3.6$~s the conversion of
$v_i$ in km/h is $v_i=(d_i/T)(3600/1000)=d_i$.
\bibitem{com1} Similar hysteresis curves are obtained when considering a spread-out
car-size distribution that follows a uniform random law
with mean value $d_{car}=5$ m.
\end{references}
\end{document}